\documentclass{aa}
\usepackage{graphicx,color}
\usepackage{txfonts}

\usepackage{color}

\begin{document}

\title{Structure and kinematics of the interacting group NGC 5098/5096\thanks{Reduced optical images in $g$ and $r$ bands are available in electronic form at the CDS via anonymous ftp to cdsarc.u-strasbg.fr (130.79.128.5).}}

\author{G.~B.~Lima Neto\inst{1}
    \and H.~V. Capelato\inst{2}
    \and F. Durret\inst{3}
    \and R.~E.~G. Machado\inst{4}
        }

\institute{Instituto de Astronomia, Geof\'isica e Ci\^encias Atmosf\'ericas, Universidade de S\~ao Paulo, Rua do Mat\~ao 1226, S\~ao Paulo/SP, Brazil\\
\email{gastao@astro.iag.usp.br}
         \and
    N\'ucleo de Astrof\'\i sica Te\'orica (NAT), Universidade Cidade de S\~ao Paulo, Rua Galv\~ao Bueno, 868, 01506-000, São Paulo, Brazil
        \and
    Sorbonne Universit\'e, CNRS, UMR~7095, Institut d'Astrophysique de Paris, 98bis Bd Arago, 75014, Paris, France
        \and
    Departamento Acad\^emico de F\'isica, Universidade Tecnol\'ogica Federal do Paran\'a, Av. Sete de Setembro 3165, Curitiba, Brazil
        }

\date{Received ??????? ??, 20??; accepted March 20, 2025}

\abstract
%context
{Most galaxies in the Universe are found in groups, which have various morphologies and dynamical states. Studying how groups evolve is an important step for our understanding in both large-scale structure formation and galaxy evolution.}
{We analysed the system composed by two groups at $z \simeq 0.037$, NGC~5098, a group dominated by a pair of elliptical galaxies, and NGC~5096, a compact system which appears to be interacting with NGC~5098. We aim to describe its current dynamical state in order to investigate how it fits in our current cosmological framework.}
{Our analysis is based on deep Canada-France-Hawaii Telescope (CFHT/MegaCam) $g$ and $r$ imaging, archival \textit{Chandra} X-ray data, and publicly available data of the galaxy redshift distribution. We model the surface brightness of the 12 brightest galaxies in the field-of-view and investigate the diffuse intragroup light that we detect. With a redshift sample of 112 galaxies, we study the dynamical states of both groups.}
{We detect low surface brightness diffuse light associated with both galaxy-galaxy interactions and a possible group-group collision. The substructure we found in velocity space indicates a past interaction between both groups. This is further corroborated by the X-ray analysis.}
{We conclude that NGC~5098 and NGC~5096 form a complex system, that may have collided in the past, producing a sloshing observed in X-rays and a large scale diffuse component of intragroup light as well as some important tidal debris.}


\titlerunning{Structure and dynamical history of the merging group NGC 5098}
\authorrunning{Lima Neto et al.}
\maketitle

\section{Introduction}

Groups of galaxies, collapsed structures in the mass range of a few $10^{12}\,M_\odot$ up to $\sim 10^{14}\,M_\odot$, are objects that host more than half of all galaxies in the present Universe \citep[e.g.,][]{Eke2004,Robotham2011}. Galaxy groups are important building blocks for the formation and growth of galaxy clusters, and also play a significant role on the evolution of the galactic population of cluster members through a pre-processing activity, i.e., the evolution of galaxies in groups before these groups fall into and merge with clusters \citep[e.g.,][]{Fujita2004,Bianconi2018,Han2018}. Furthermore, galaxy groups have many different configurations, such as loose, compact, rich, poor, and fossil \citep[e.g.,][]{Lovisari2021}.

Galaxies in groups are prone to interact with each other and with the intragroup medium. Since the velocity dispersion in groups is usually a few hundred km/s, comparable to the stellar velocity dispersion in large galaxies, galactic encounters will have enough time to produce tidal forces that will act upon galaxy members. Moreover, galactic collisions in low velocity dispersion systems will often end in mergers after a few gigayears \citep[e.g.,][]{Binney2008,Jiang2008,Solanes2018}.

An important consequence of tidal interactions in groups (and also clusters) is the build up of a diffuse stellar component \citep[see, e.g.,][]{Contini2021,Montes2022,Arnaboldi2022}. This component is made of stars ripped away from galaxies during these interactions. For simplicity, we call this component the intracluster light, ICL. The material stripped from galaxies spreads into the host group or cluster, relaxing into the gravitational potential well \citep[e.g.,][]{Murante2007,Rudick2009,Contini2014,Burke2015,Contini2023,Joo2023}. The observation of ICL can tell us how this process acts and give us clues about the dynamical history of the host halo. However, this is not completely understood yet \cite[e.g.,][]{Willman2004,Jimenez2023,Ragusa2023,Brough2024,Contini2024}. The ICL is a relic of past galactic interactions, both between galaxies themselves, and between galaxies and the cluster gravitational potential, as well as of the ram pressure stripping process \citep[see recent reviews, e.g.,][]{Montes2022,Contini2021,Contini2024}.

The ICL is often defined as the stars that are not bound to any particular galaxy member in a group or cluster, but respond to the general gravitational potential, as first pointed out by \citet{Zwicky1937,Zwicky1951} \citep[see also,][]{Willman2004,Rudick2011,Jimenez2016,Ko2018,Contini2024}. This theoretical definition is, however, difficult to use observationally, since we cannot measure directly the bound state of intracluster stars forming the ICL.

The above definition of ICL can in principle be used with $N$-body simulations. For instance, \citet{Sampaio2021}, based on the Illustris TNG300 simulations \citep[see][for details of the TNG simulations]{Pillepich2018}, conclude that the diffuse ICL is not a good tracer of the total mass density in clusters, even though it is formed by stars orbiting the cluster global potential. However, from an observational standpoint, \citet{Montes2019} find that the ICL does trace the total mass determined through weak-lensing analysis. Based on the Cluster-Eagle simulation, \citet{Alonso2020} find that the shape of the stellar distribution closely follows the total mass distribution, but with a different radial profile. More recently, \citet{Diego2023} analysed the cluster SMACS0723, observed by JWST, concluding that the ICL has a steeper radial profile than the weak-lensing derived mass distribution. Moreover, the ICL has features that have no obvious counterpart with the total cluster mass.

Some authors use a pragmatic approach, defining the ICL as the diffuse component of stars that has a surface brightness below a certain threshold. For instance, \citet{Feldmeier2004} use surface brightness cuts between 26.0 and 27.5~mag~arcsec$^{-2}$ to estimate the ICL contribution in four Abell clusters without a central cD galaxy. \citet{Furnell2021}, on the other hand, use a surface brightness threshold of 25~mag~arcsec$^{-2}$ on a sample of 18 clusters observed by XMM-\textit{Newton} between redshifts 0.1 and 0.4. This is the simplest approach for defining the ICL.

 Other authors define the ICL as the residual left after subtracting a light distribution model from the galactic images in a group or cluster. \citet{Gonzalez2005} used two-component models on 24 clusters with a dominant cD galaxy, and interpreted the more luminous, extended component as the ICL linked to the global cluster gravitational potential. \citet{Durret2019} also use two-component model fits with HST images of clusters dominated by a central bright galaxy to analyse the ICL in the redshift range [0.2-0.9]. A detailed discussion of different methods of ICL definition is presented by \citet{Contini2021}.

In this work, we describe a photometric and kinematic analysis of the galaxy group dominated by the pair of elliptical galaxies NGC~5098a and NGC~5098b, and its neighbouring group centred on NCG~5096. This complex system shows features that suggest a recent dynamical interaction that we will explore here and describe below.

This paper is organised as follows. Section~\ref{sec:NGC5098} describes the NGC~5098 group and previous results. In section~\ref{sec:X-ray} we describe the \textit{Chandra} X-ray data that we use and in section~\ref{sec:optical} we analyse the CFHT optical images. We then discuss the diffuse light detected in section~\ref{sec:ICL}. In section~\ref{sec:Dynamic} we analyse the group dynamics based of the available redshifts. Finally, in section~\ref{sec:Discussion} we discuss our results and conclude in Sec.~\ref{sec:Conclusion}.
In this paper, assuming the following cosmological parameters, $H_0 = 70\,$km~s$^{-1}$~Mpc$^{-1}$, $\Omega_M = 0.3$, $\Omega_\Lambda = 0.7$, the plate-scale is $43 h_{70}^{-1}\,$kpc/arcmin and the luminosity distance is $171 h_{70}^{-1}\,$Mpc.

\section{The NGC 5098/5096 group}
\label{sec:NGC5098}

NGC 5098, discovered by John Herschel in 1827, is actually a pair of early-type galaxies, the brightest ones in the RGH~80 group \citep{Ramella1989}, also known as the NGC~5098 group. NGC 5098a is the galaxy to the west, while NGC~5098b is to the east. \citet{Randall2009} computed the absolute magnitudes for both galaxies based on the SDSS\footnote{\texttt{https://skyserver.sdss.org/dr18}} data: $M_B = -21.131$ and $M_V = -22.097$ for NGC~5098a and $M_B = -20.845$ and $M_V = -21.770$ for NGC~5098b. Their redshifts are $z = 0.03606 \pm 0.00001$ for NGC~5098a \citep{Lee2017} and $z = 0.03744 \pm 0.00008$ for NGC~5098b \citep{Miller2002}.

\begin{figure}
\includegraphics[width=\columnwidth]{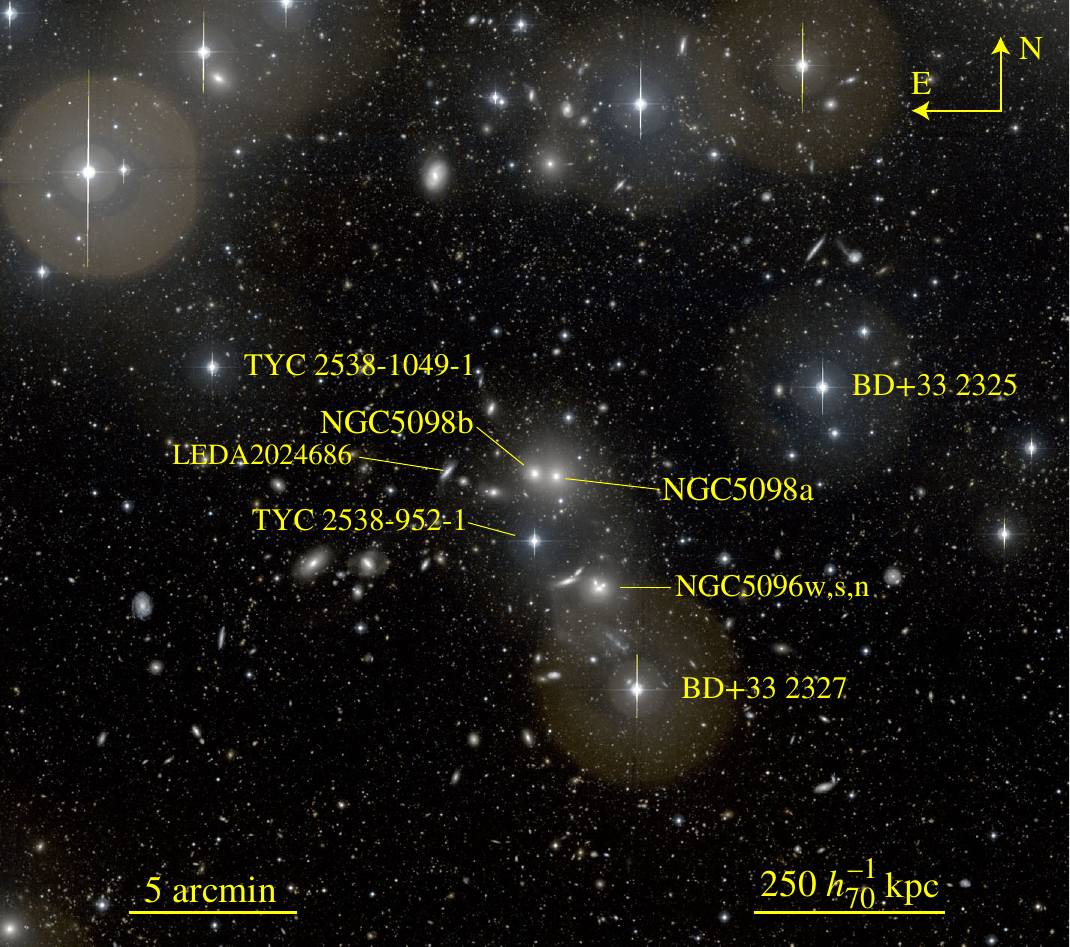}
\caption{Colour image of the NGC~5098 Group based on CFHT/MegaCam images in the $g$ and $r$ bands. We show the galaxy pair NGC~5098a and b, and the three brightest stars in the field of view of the group core. NGC~5096 forms a subgroup with three galaxies: w (west), s (south) and n (north)  The brightness is scaled logarithmically. LEDA2024686 is a foreground galaxy at redshift $z = 0.01819$. North is top and east is left, as in all figures.}
\label{fig:NGC5098_CFHT_zoomout}
\end{figure}

% Radio observation and spectra SDSS
One of the two main galaxies of the NGC~5098 group has an extended radio source, 1317+33, associated to NGC~5098a, detected at 1415, 4850 and 4996~MHz \citep{Fanti1977,Becker1991}. Moreover, the SDSS spectrum of this galaxy shows a broad H$\alpha$ line as a some [O\textsc{iii}] and [S\textsc{ii}] emission lines.
This suggests the presence of a moderately active galactic nucleus in this galaxy.

The NGC~5098 group was extensively observed in X-rays, detected in the ROSAT All-Sky Survey (RASS) as an extended source with luminosity in the [0.1--2.4 keV] band $L_X = 3.33 \times 10^{43} h_{50}^{-2}\,$erg~s$^{-1}$ within $1.0\,h_{50}^{-1}\,$Mpc \citep{Mahdavi2000} and was included in the Brightest Cluster Sample as source RXJ1320.1+3308 \citep{Ebeling1998}. 

\cite{Xue2004} analyzed XMM-\textit{Newton} data, detecting the X-ray emission up to $462\,h_{50}^{-1}\,$kpc, with a mean temperature of $1.01 \pm 0.01$~keV and a peak temperature of 1.3~keV at $0.11\, R_{200}$, with a cool-core in the centre at 0.83~keV, and a gradual drop  outward. The peak of the X-ray emission coincides spatially with NGC~5098a.

\cite{Randall2009} analyzed a deep \textit{Chandra} observation made with the ACIS-S3 detector. Thanks to the high spatial resolution, they discovered the presence of a spiral-like arm due to gas sloshing in the centre, starting from the position of NGC~5098a, and unwinding outwards to the north and west. They argue that it is probably the galaxy NGC~5098b which is the perturber causing the sloshing. They also identify ``bubbles'' (regions with a decrease in the X-ray emission) associated with the radio maps from VLA observations at 1.45~GHz.

Towards the south, about $3.6^\prime$ from NGC~5098a and b, there is a substructure dominated by the triple system of NGC~5096 (with the letters s, w, and n, for their relative positions, south, west and north, see Fig.~\ref{fig:NGC5098_CFHT_zoomout}), that is considered a part of the NGC~5098 group \citep[e.g.,][]{Ramella1995}.

NGC~5096 was classified as a Hickson-like compact group by \citet[][their group number 826]{Zandivarez2022}. The three central galaxies, NGC~5096s, n, and w (also shown in Fig.~\ref{fig:NGC5098_CFHT_zoomout}) have a median redshift $z=0.0394$ and a velocity dispersion of only 44~km~s$^{-1}$.

\section{X-ray data}
\label{sec:X-ray}

NGC~5098 was extensively studied in X-rays by \citet{Xue2004}, \citet{Mahdavi2005} and \citet{Randall2009}, as mentioned above. In order to compare the X-ray emission, which mostly traces the intragroup hot plasma distribution, with our optical images (described below), we have downloaded the three \textit{Chandra} publicly available exposures with ObsID: 2231 (PI Fukazawa, 11~ks), 3352 (PI Ueda 2.66~ks), and 6941 (PI Buote, 39.13~ks).

We have reprocessed these exposures adopting the standard procedure with the software package \textsc{ciao}~4.14 and \textsc{caldb}~4.10.2 from the \textit{Chandra X-ray Center} (CXC)\footnote{\texttt{https://cxc.harvard.edu/ciao/}} using the script \texttt{chandra\_repro}\footnote{\texttt{https://cxc.cfa.harvard.edu/ciao/ahelp/\\ chandra\_repro.html}} 
with the default parameters. We then combined the event files and produced a single broad band (0.5--7.0 keV) exposure-map corrected image with the script \texttt{merge\_obs}\footnote{\texttt{https://cxc.cfa.harvard.edu/ciao/ahelp/merge\_obs.html}}, binning the observation with a factor of 4, resulting in a scale of 1.968~arcsec per pixel.

\begin{figure}[htb]
    \centering
    \includegraphics[width=\columnwidth]{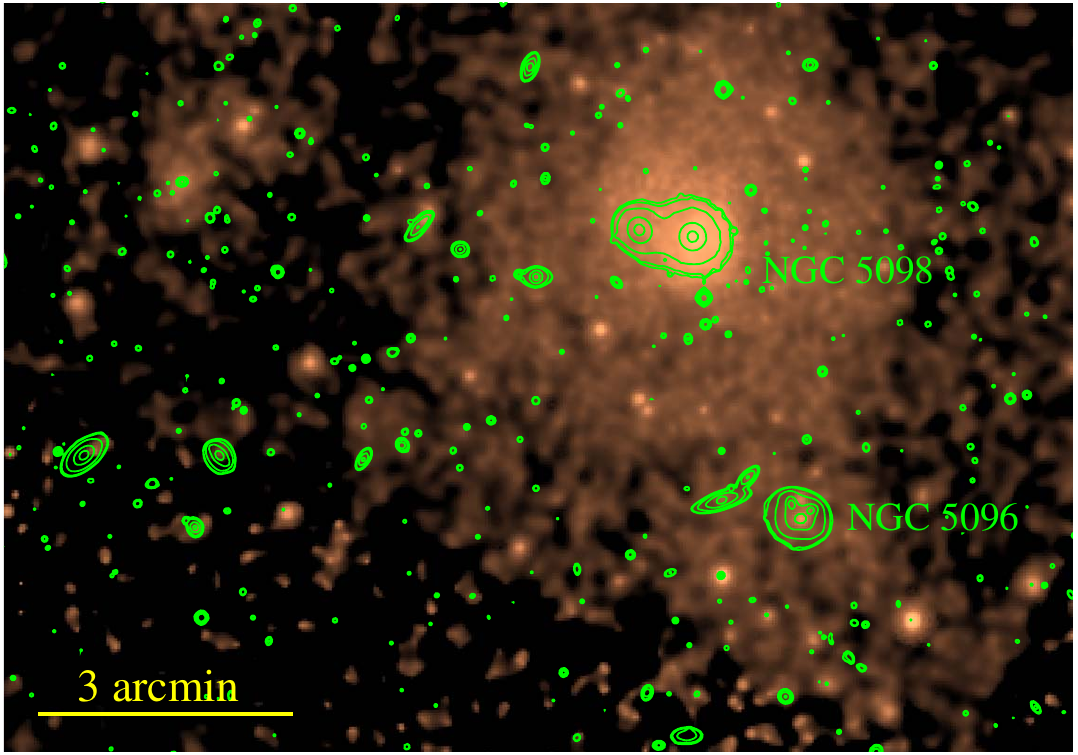}
    \caption{Exposure-map corrected image of the combined \textit{Chandra} exposures in logarithmic scale. Green contours are isopleths from the $r$ band image (without the bright stars), ranging logarithmically from 600 to 2300 ADU (21.4 to 19.8 mag/arcsec$^2$, see Sec.~\ref{sec:optical}). 3~ arcmin correspond to $147 h_{70}^{-1}\,$kpc.}
    \label{fig:Chandra_contGalaxies}
\end{figure}

Figure~\ref{fig:Chandra_contGalaxies} shows the broad band \textit{Chandra} image smoothed with an adaptive kernel using the tool \texttt{dmimgadapt} from \textsc{ciao} with the $r$ band isopleths superposed. The main X-ray emission comes from an extended halo centred on NGC~5098a, while a secondary extended, diffuse emission comes from the NGC~5096 substructure. The diffuse emission on both sources appears connected, showing some asymmetry, an excess westward from NGC~5098 and another X-ray excess eastward from NGC~5096. This is suggestive of a physical connection between both groups.

The sloshing arm detected by \citet{Randall2009} can be seen as a faint X-ray excess to the north of NGC~5098b, spiralling outward clockwise. \citeauthor{Randall2009} suggested that this feature could be due to the arrival and gravitational interaction of NGC~5098b with NGC~5098a, the main galaxy of the group.

\section{Optical analysis}
\label{sec:optical}

The NGC~5098 group was observed with MegaCam at the 3.6m Canada-France-Hawaii Telescope (CFHT, observation Program 13AF002 in May and June 2013) in the $g$ and $r$ bands. We have applied the Elixir-LSB pipeline for the image reduction, which was designed for dealing with very faint extended sources \citep{Duc2011}. The images in both bands were further binned, resulting in a plate scale of $0.561^{\prime\prime}$ per pixel. The total effective exposure time was 2240~s (seven individual 320~s exposures) for each band. Figure~\ref{fig:NGC5098_CFHT_zoomout} shows a $\sim 0.5$~deg$^2$ region centred at $13^{\rm h}20^{\rm m}17.74^{\rm s}$, $+33^\circ08^\prime43.6^{\prime\prime}$ (J2000). On this image, it is possible to see the common stellar envelope around the NGC~5098 pair and another stellar envelope around the NGC~5096 triplet.

Both images, in the $g$ and $r$ bands, have a zero-point ZP = 29.61~mag. Their magnitudes are obtained with $m = 29.61 - 2.5 \log(\mbox{ADU}/0.3147)$, where the constant 0.3147 corresponds to the plate scale of $0.561^{\prime\prime}$ per pixel and ADU is ``Analog-Digital Units'' used on these images.

We checked that the images in both bands were indeed flat throughout the whole field of view by applying a wavelet decomposition%
\footnote{We have applied the code \texttt{wvdecomp} from the package \textsc{zhtools}, \texttt{https://github.com/avikhlinin/wvdecomp}, \citep[see,][]{Vikhlinin1997}.} following a technique similar to that employed by \citet{Guennou2012} and \citet{Ellien2021}.
The background levels, measured using a 3$\sigma$-clipping method around the median value, are $532.8\pm 4.2$ and $327.8\pm2.8$ ADU per pixel for the $r$ and $g$ bands, respectively. In magnitude units, this corresponds to $21.538 \pm 0.009$ and $22.066 \pm 0.010$~mag/arcsec$^2$ for the $r$ and $g$ bands, respectively.

\subsection{Scattered light from stars}

Although the deep images obtained with CFHT/MegaCam and processed by the Elixir-LSB pipeline have a very good flat-field, the bright point-like sources show a circular halo of scattered light, due to telescope optics and multiple internal reflections. 
\citet{Karabal2017} proposed a method for dealing with these ``ghost'' images of scattered light on images acquired with MegaCam. They model the bright star halos with an empirical model that is then subtracted from the original image.

In order to analyse the very faint extended emission, we have either to mask or subtract these stellar halos. We applied a method similar to \citet{Karabal2017}, modelling the stellar scattered light with  the \texttt{galfit} software \citep{Peng2002,Peng2010}. This program is flexible enough that we can model each star independently with a combination of elliptically symmetric models.

\begin{figure*}[htb]
    \centering
    \includegraphics[width=0.9\textwidth]{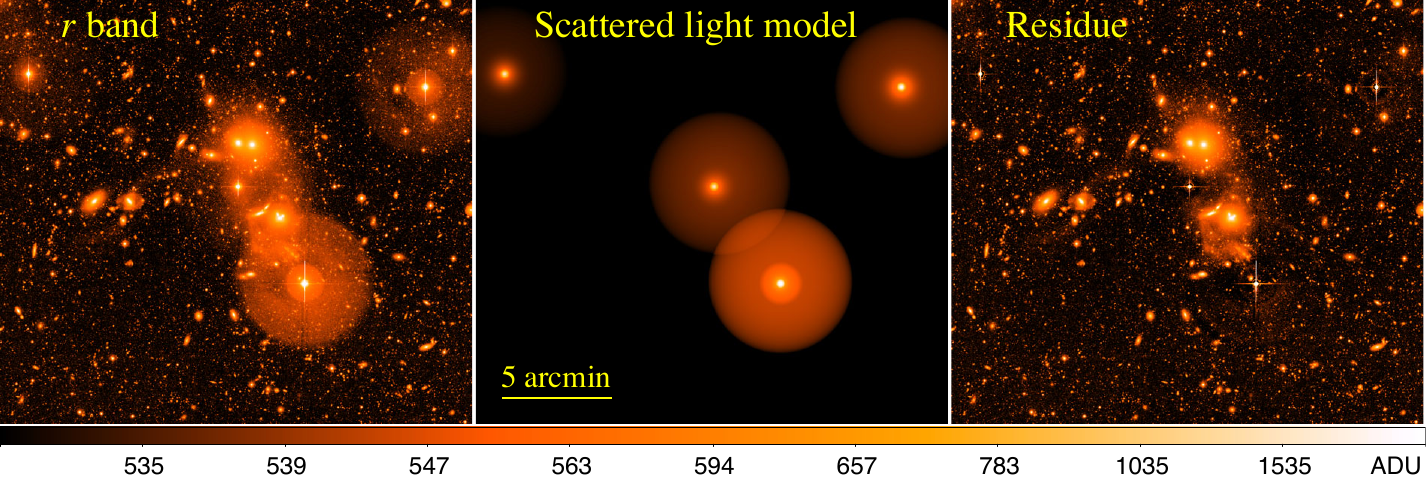}
    \caption{Illustration of the star subtraction. \textsf{Left:} original $r$ band image. \textsf{Center:} Star models using three Ferrers profiles (see text for more details). \textsf{Right:} Residual image after subtracting the star models from the $r$ band image. The colour scale is the same in all panels. For reference, 535 ADU is 21.53 mag and 1535 ADU is 20.39 mag.}
    \label{fig:estrelas_r}
\end{figure*}

We have adopted a modified Ferrers model \citep{Ferrers1877,Laurikainen2005,Peng2010}, which has a very flat core and an outer region that is truncated. We found that its functional form can be used to build an empirical description of the bright star light forming ``ghosts''. The modified Ferrers surface brightness radial profile is:
\begin{equation}
I(R) = I_0 \left( 1 - (R/R_{\rm out})^{2-\beta} \right)^\alpha \, ,
\end{equation}
that is only defined for $R \le R_{\rm out}$. The parameters $\alpha$ and $\beta$ define the shape of this profile and are kept free in the fitting. We have used a combination of three Ferrers profiles for each of the four brightest stars near the centre of the NGC~5098 group (see Table~\ref{tab:estrelas}). Figure~\ref{fig:estrelas_r} shows the result of modelling four stars and subtracting them from the MegaCam $r$ band image. The four stars are fitted simultaneously. The inner saturated part of the ghost scattered light, inside a radius of 12~arcsec centred on each bright star, is not modelled and we mask it in further analysis.

\begin{table}[htb]
\caption{The brightest stars near the field-of-view of NGC~5098 and NGC~5096.}
\label{tab:estrelas}
 \setlength{\tabcolsep}{0.9ex}\small
\begin{tabular}{lcccc}
\hline
     Star name       &  RA (J2000)    & Dec (J2000)  & BTmag & VTmag \\
\hline
     BD+33 2325      & 199.9042067 & 33.1874902 & 11.43 & 10.83 \\
     BD+33 2327      & 200.0138959 & 33.0381647 & 11.89 & 10.45 \\
     TYC 2538-952-1  & 200.0741837 & 33.1115314 & 12.01 & 11.55 \\
     TYC 2538-1049-1 & 200.2642504 & 33.1972697 & 12.31 & 11.76 \\
\hline
     & 
\end{tabular}
\tablefoot{This are the stars that were modelled and subtracted from our MegaCam images. Data from the Tycho-2 catalogue \citep{Hog2000}. BTmag and VTmag are Tycho magnitudes in the $B$ and $V$ bands, respectively.}
\end{table}

\subsection{Modelling the brightest galaxies}

We have also used \texttt{galfit} to model the brightest galaxies in the NGC~5098 group using a combination of S\'ersic profiles \citep{Sersic1963} with a flat background, and taking into account the point spread function (PSF), using the star subtracted images in the $g$ and $r$ bands. 

We have chosen 20 non-saturated stars, far from bright extended objects, and stacked them in order to estimate the PSF. This was done by 2D fitting a  Moffat radial profile described by Eq.~(\ref{eq:Moffat}):
\begin{equation}
\mbox{PSF}(R) = \frac{\beta-1}{\pi\, \alpha} \left[ 1 + (R/\alpha)^2 \right]^{-\beta}\, ,
~\mbox{with FWHM} = \alpha\, 2\sqrt{2^{1/\beta} -1}\, ,
\label{eq:Moffat}
\end{equation}
where FWHM is the full width at half maximum related to the Moffat parameters $\alpha$ and $\beta$. With the 2D-fit, we produced a synthetic image representing the PSF that we can use with \texttt{galfit}.

\begin{figure}[htb]
    \centering
    \includegraphics[width=0.96\columnwidth]{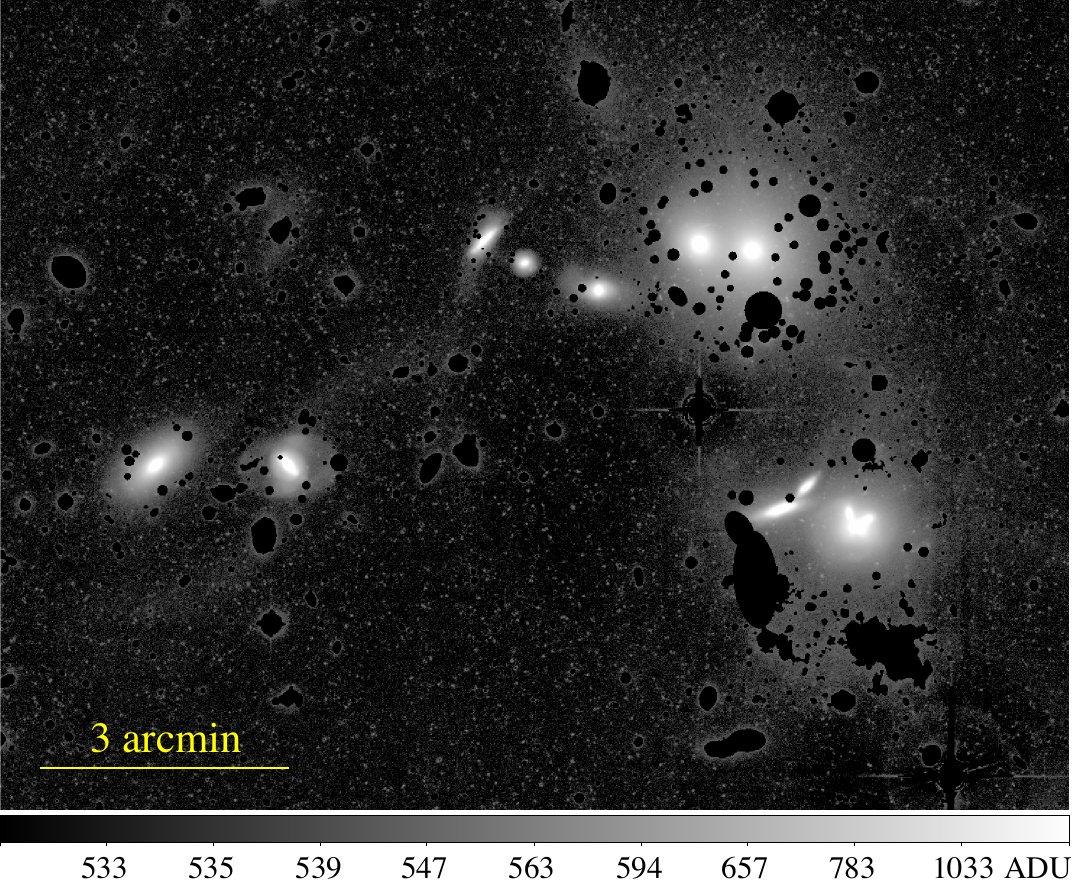}
    \caption{Masked $r$ band image, with bright stars subtracted, that was used to fit the twelve bright galaxies with \texttt{galfit}. The gray scale is logarithmic. Notice the large scale low surface brightness diffuse light to the west.}
    \label{fig:NGC5098_r_semStar_masked}
\end{figure}

We then produced an image mask that is used to fit the galaxies of interest without contamination by other sources. For the $r$ band, we have masked every source brighter than 544 ADU (21.52 mag/arcsec$^2$), which is about $2.7 \sigma$ above the background. For the $g$ band, we used a threshold of 335 ADU (22.0 mag/arcsec$^2$), about $2.6 \sigma$ above the background. The masked $r$ band image is shown in Fig.~\ref{fig:NGC5098_r_semStar_masked}. With this choice of threshold we mask almost all point sources and most of the extended smaller objects, but still leaving unmasked the intragroup light and extended stellar components of the brightest galaxies.

We chose to model the twelve brightest galaxies in the central region of NGC~5098, including the NGC~5096 substructure (Table~\ref{tab:Resu12galaxies}). These galaxies include the two main elliptical galaxies in the core of NGC~5098 and two other galaxies eastward. The third galaxy to the east, LEDA2024686, is actually a foreground galaxy but we also modelled it since it could contribute to the surface brightness of the extended envelope of the two main elliptical galaxies.

\begin{table*}[htb]
\caption{The twelve galaxies that were modelled with \texttt{galfit}.}
\label{tab:Resu12galaxies}
\begin{tabular}{cc c cccl}
\hline
RA (J2000)   & Decl (J2000) & SDSS ID             & gmag                & rmag              & redshift  & Name \\
\hline
200.06136849 & +33.14341078 & J132014.72+330836.2 & 14.306 $\pm$ 0.002 & 13.468 $\pm$ 0.002 &  0.03606   & NGC5098a \\
200.07391197 & +33.14474057 & J132017.74+330841.0 & 14.608 $\pm$ 0.002 & 13.816 $\pm$ 0.002 &  0.037436* & NGC5098b \\
200.09814104 & +33.13550954 & J132023.55+330807.8 & 16.200 $\pm$ 0.003 & 15.368 $\pm$ 0.003 &  0.03724   & LEDA2024371 \\
200.11595638 & +33.14094451 & J132027.82+330827.3 & 17.326 $\pm$ 0.006 & 16.522 $\pm$ 0.004 &  0.035808* & LEDA2024547 \\
\hline
200.17245095 & +33.10038956 & J132041.38+330601.3 & 15.407 $\pm$ 0.003 & 14.660 $\pm$ 0.002 &  0.03606   & LEDA2023331 \\
200.20426174 & +33.10039187 & J132049.02+330601.3 & 15.293 $\pm$ 0.002 & 14.488 $\pm$ 0.002 &  0.03458   & LEDA2023332 \\
\hline
200.03621197 & +33.08795822 & J132008.69+330516.6 & 15.058 $\pm$ 0.002 & 14.057 $\pm$ 0.002 &  0.03925   & NGC5096s \\
200.03816676 & +33.09077346 & J132009.16+330526.7 & 16.141 $\pm$ 0.003 & 15.370 $\pm$ 0.003 &  0.039404* & NGC5096n \\
200.03368592 & +33.08957994 & J132008.08+330522.4 & 16.374 $\pm$ 0.004 & 15.612 $\pm$ 0.003 &  0.039611* & NGC5096w \\
200.05459134 & +33.09162369 & J132013.10+330529.8 & 15.540 $\pm$ 0.003 & 14.684 $\pm$ 0.002 &  0.03916   & LEDA2023056 \\
200.04847044 & +33.09609510 & J132011.63+330545.9 & 17.150 $\pm$ 0.005 & 16.300 $\pm$ 0.004 &  0.039614* & LEDA2023215 \\
\hline
200.12590289 & +33.14523506 & J132030.21+330842.8 & 16.019 $\pm$ 0.003 & 15.805 $\pm$ 0.004 &  0.01819   & LEDA2024686 \\
\hline
\end{tabular}
\tablefoot{The first four lines are the NGC~5098 group, the following 
two lines are the galaxy pair to the east, the next five lines are the galaxies in the NGC~5096 substructure. 
The last line is a foreground galaxy that was also modelled. Data are from SDSS DR18, except for the redshifts
marked with *, that are from LEDA. The AB magnitudes in the $g$ and $r$ bands are from the SDSS, gmag and rmag, respectively.}
\end{table*}

Further east, we also modelled two possibly spiral galaxies that, as we discovered, contribute to the intragroup light. Finally, there are five galaxies in a very compact configuration, forming the central substructure of NGC~5096, that were also modelled.

\begin{figure*}
    \centering
    \includegraphics[width=0.98\textwidth]{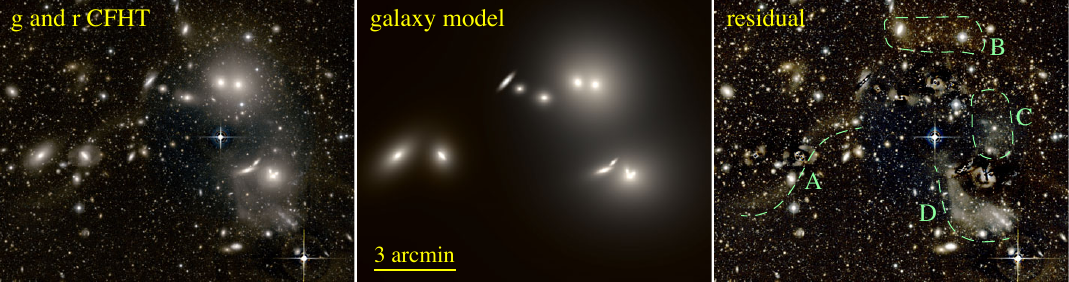}
    \caption{\textsf{Left}: ``true colour'' RGB image using $g$, $r$ and the the mean value of $g$ and $r$ bands. \textsf{Middle}: galaxies modelled with \texttt{galfit}, also in ``true colour''. \textsf{Right}: residual of the CFHT image minus the model image, showing the detected diffuse light (see text for more details). All images are in logarithmic scale.}
    \label{fig:modelGals_residue}
\end{figure*}

\begin{table*}[htb]
    \centering
    \caption{Best fit results for the $g$ band.}
    \label{tab:galfit_result_g}
    \begin{tabular}{lcccccc}
\hline
 name      &       mag\_g   & $R_{\rm eff}$ [arcsec] &      $n$          & $b/a$   & PA & $L\; [10^9 L_\odot]$\\
\hline
NGC5098a   & 15.3937 $\pm$ 0.0086 &  3.64 $\pm$ 0.02 &  2.116 $\pm$ 0.007 &  1.00 &   40.18 &  22.3 \\
NGC5098a   & 13.7793 $\pm$ 0.0038 & 52.19 $\pm$ 0.15 &  2.677 $\pm$ 0.021 &  0.92 &   28.33 &  98.4 \\
\\[-1.5ex]
NGC5098b   & 14.5652 $\pm$ 0.0008 &  6.67 $\pm$ 0.01 &  3.248 $\pm$ 0.005 &  0.89 &   34.71 &  47.7 \\
\\[-1.5ex]
LEDA2024371& 15.9482 $\pm$ 0.0011 &  3.16 $\pm$ 0.01 &  3.758 $\pm$ 0.012 &  0.81 &   88.22 &  13.4 \\
\\[-1.5ex]
LEDA2024547& 17.2384 $\pm$ 0.0020 &  2.61 $\pm$ 0.01 &  2.553 $\pm$ 0.014 &  0.82 &  -83.27 &  4.07 \\
\hline	
LEDA2023331& 16.9627 $\pm$ 0.0032 &  6.37 $\pm$ 0.01 &  0.244 $\pm$ 0.002 &  0.25 &   42.79 &  5.25 \\
LEDA2023331& 15.5242 $\pm$ 0.0010 &  4.92 $\pm$ 0.01 &  3.385 $\pm$ 0.007 &  0.82 &   27.55 &  19.7 \\
\\[-1.5ex]
LEDA2023332& 15.0016 $\pm$ 0.0007 &  6.62 $\pm$ 0.01 &  4.502 $\pm$ 0.008 &  0.54 &  -49.60 &  31.9 \\
\hline	
NGC5096s   & 13.7989 $\pm$ 0.0025 & 37.95 $\pm$ 0.25 &  9.201 $\pm$ 0.023 &  0.81 &   88.57 &  96.7 \\
\\[-1.5ex]
NGC5096n   & 15.8877 $\pm$ 0.0016 &  2.14 $\pm$ 0.01 &  4.563 $\pm$ 0.024 &  0.56 &   12.42 &  14.1 \\
\\[-1.5ex]
NGC5096w   & 16.9535 $\pm$ 0.0076 &  0.98 $\pm$ 0.01 &  2.553 $\pm$ 0.047 &  0.64 &  -62.69 &  5.29 \\
NGC5096w   & 17.2145 $\pm$ 0.0091 &  3.89 $\pm$ 0.02 &  0.797 $\pm$ 0.008 &  0.90 &  -49.68 &  4.16 \\
\\[-1.5ex]
LEDA2023056& 15.6807 $\pm$ 0.0018 &  6.16 $\pm$ 0.03 &  5.212 $\pm$ 0.028 &  0.45 &  -69.69 &  17.1 \\
LEDA2023056& 16.6541 $\pm$ 0.0038 &  8.50 $\pm$ 0.01 &  0.479 $\pm$ 0.003 &  0.24 &  -71.84 &  6.97 \\
\\[-1.5ex]
LEDA2023215& 16.9351 $\pm$ 0.0017 &  4.52 $\pm$ 0.01 &  0.896 $\pm$ 0.004 &  0.36 &  -46.28 &  5.38 \\
\hline	
LEDA2024686& 15.9819 $\pm$ 0.0011 &  6.18 $\pm$ 0.01 &  0.902 $\pm$ 0.002 &  0.28 &  -42.70 &  3.23 \\
\hline
\end{tabular}
\tablefoot{Two consecutive lines mean that a double S\'ersic model was used. The galaxies follow the same order as in Table~\ref{tab:Resu12galaxies}. The total magnitude for each model is mag\_g, $R_{\rm eff}$ is the effective radius, $n$ is the S\'ersic index, $b/a$ is the axial ratio, and PA is the position angle measured from north to east. Error bars are $1 \sigma$ confidence level. For LEDA2024686 we assume a luminosity distance of 85~Mpc.}
\end{table*}

\begin{table*}[htb]
    \centering
    \caption{Same as Table~\ref{tab:galfit_result_g}, but for the $r$ band.}
    \label{tab:galfit_result_r}
    \begin{tabular}{lcccccc}
\hline
 name       &       mag\_r          & $R_{\rm eff}$ [arcsec] &   $n$    &   $b/a$ &  PA & $L\; [10^9 L_\odot]$\\
\hline
NGC5098a    &  14.5624 $\pm$ 0.0094 &  3.63 $\pm$ 0.03 & 2.229 $\pm$ 0.011 &  1.00 &  40.18 & 31.3 \\
NGC5098a    &  13.5287 $\pm$ 0.0037 & 36.40 $\pm$ 0.15 & 1.954 $\pm$ 0.013 &  0.91 &  28.33 & 81.2 \\
\\[-1.5ex]
NGC5098b    &  13.8951 $\pm$ 0.0006 &  6.59 $\pm$ 0.01 & 3.492 $\pm$ 0.004 &  0.88 &  34.71 & 57.9 \\
\\[-1.5ex]
LEDA2024371 &  15.2893 $\pm$ 0.0010 &  2.64 $\pm$ 0.01 & 3.703 $\pm$ 0.011 &  0.81 &  88.22 & 16.0 \\
\\[-1.5ex]
LEDA2024547 &  16.6081 $\pm$ 0.0020 &  2.29 $\pm$ 0.01 & 2.381 $\pm$ 0.013 &  0.82 & -83.27 & 4.76 \\
\\
LEDA2023331 &  16.0467 $\pm$ 0.0026 &  5.39 $\pm$ 0.01 & 0.440 $\pm$ 0.002 &  0.29 &  42.65 & 7.98 \\
LEDA2023331 &  14.9481 $\pm$ 0.0011 &  4.87 $\pm$ 0.01 & 4.161 $\pm$ 0.011 &  0.86 &  14.82 & 22.0 \\
\\[-1.5ex]
LEDA2023332 &  14.3228 $\pm$ 0.0007 &  6.12 $\pm$ 0.01 & 4.804 $\pm$ 0.008 &  0.54 & -49.60 & 39.1 \\
\hline	
\\[-1.5ex]
NGC5096s    & 13.4048 $\pm$ 0.0012  & 14.80 $\pm$ 0.05 & 7.479 $\pm$ 0.013 &  0.81 &  88.57 & 91.0 \\
\\[-1.5ex]
NGC5096n    & 15.2117 $\pm$ 0.0014  &  1.99 $\pm$ 0.01 & 5.353 $\pm$ 0.025 &  0.56 &  12.42 & 17.2 \\
\\[-1.5ex]
NGC5096w    & 16.2079 $\pm$ 0.0127  &  0.95 $\pm$ 0.02 & 3.064 $\pm$ 0.045 &  0.64 & -62.69 & 6.88 \\
NGC5096w    & 16.8023 $\pm$ 0.0197  &  3.54 $\pm$ 0.03 & 0.850 $\pm$ 0.013 &  0.90 & -49.68 & 3.98 \\
\\[-1.5ex]
LEDA2023056 & 15.0609 $\pm$ 0.0030  &  4.53 $\pm$ 0.02 & 4.523 $\pm$ 0.018 &  0.45 & -70.04 & 19.8 \\
LEDA2023056 & 15.8953 $\pm$ 0.0068  &  8.67 $\pm$ 0.03 & 0.524 $\pm$ 0.004 &  0.24 & -71.90 & 9.18 \\
\\[-1.5ex]
LEDA2023215 & 16.2224 $\pm$ 0.0017  &  4.19 $\pm$ 0.01 & 0.988 $\pm$ 0.005 &  0.36 & -46.28 & 6.79 \\
\hline	
LEDA2024686 & 15.7626 $\pm$ 0.0014  &  6.21 $\pm$ 0.01 & 0.969 $\pm$ 0.003 &  0.28 & -42.70 & 2.59 \\
\hline
    \end{tabular}
\end{table*}

We applied the latest version of the \texttt{galfit} software to model the above galaxies using either one or two S\'ersic models. Figure~\ref{fig:modelGals_residue} shows the modelled galaxies and the residual (original image minus the model image). The resulting best-fit parameters are shown in Tables~\ref{tab:galfit_result_g} and \ref{tab:galfit_result_r}. The luminosities were calculated assuming no $k$-correction or evolution (given the low redshift of the galaxies, it is an adequate approximation) and adopting for the solar absolute magnitudes $M_{\odot, g} = 5.11$ and $M_{\odot, r} = 4.65$ \citep{Willmer2018}.

The best fit model for NGC~5098a has two S\'ersic components, including a luminous extended stellar envelope. The total apparent magnitude is 13.56 ($g$ band) and 13.17 ($r$ band). The corresponding total luminosity is $2.5 \times 10^{10}\, L_\odot$ in the $g$ band and $11.3 \times 10^{10}\, L_\odot$ in the $r$ band. Its companion, NGC~5098b, was modelled with a single S\'ersic profile, adding a second model did not improve the fit. The extended stellar envelope is thus probably associated with NGC~5098a.

Compared to the SDSS magnitudes given in Table~\ref{tab:Resu12galaxies}, our results have smaller values. For the $r$ band, our measurement is 0.3~mag brighter and for the $g$ band it is 0.7~mag brighter. The difference is probably due to our deeper images, allowing us to model the galaxy to a larger radius than the SDSS pipeline, and to our use of two S\'ersic components for NGC~5098a.

NGC~5096, which is actually a triple system, has one galaxy, NGC~5096w that is modelled by two S\'ersic models, while the other two, NGC~5096s and NGC~5096n, were fitted by a single S\'ersic each. The total magnitude of NGC~5096w is 16.3 ($g$ band) and 15.7 ($r$ band). The brightest component in NGC~5096s, which we identify as an extended stellar envelope, accounts for $\sim 80$\% of the total luminosity in the $g$ band and $\sim 76$\% in the $r$ band. 

The residual around NGC~5096 is still quite bright, but highly asymmetrical; adding more elliptically symmetric components did not improve the fitting. 
This excess has a colour $g-r = 0.7 \pm 0.3$, slightly bluer than the three central galaxies NGC~5096n, s and w. This may be due to this stellar diffuse component being recently created by the gravitational interaction of the three components of NGC~5096. The bluer colour may indicate a stellar stripping origin \citep[e.g.][]{DeMaio2018,Contini2019}.

Towards the northeast of NGC~5096, there is a pair of galaxies, LEDA2023056 and LEDA2023215 (the one to the north), that are separated by 24~arcsec ($20\,h_{70}^{-1}$\,kpc). We needed two S\'ersic profiles to model LEDA2023056, with total magnitudes 15.31 ($g$ band) and 14.65 ($r$ band), again very close to the SDSS values.

To the east of NGC~5098, there is a pair of bright galaxies, LEDA2023331 and LEDA2023332, 1.6~arcmin ($78\,h_{70}^{-1}$\,kpc) apart (the two leftmost galaxies in Fig.~\ref{fig:NGC5098_r_semStar_masked}). LEDA2023331, the galaxy west of the pair, needed two S\'ersic models to account for its brightness distribution. Its total magnitude is 15.3 in the $g$ band and 14.6 in the $r$ band, very close to the SDSS values (Tab.~\ref{tab:Resu12galaxies}). Its surface brightness radial profile suggests that it is a disk galaxy seen almost head-on, and it has a strong bar 20~arcsec long and 4~arcsec wide (16 and $3 h_{70}^{-1}$\,kpc).

\section{Diffuse light}  
\label{sec:ICL}

In the right panel of Fig.~\ref{fig:modelGals_residue}, the residual after the model subtraction, we highlight four structures with detected diffuse light, which we will describe below in more detail.

\subsection{Region A}

\begin{figure}[htb]
    \centering
    \includegraphics[width=0.96\columnwidth]{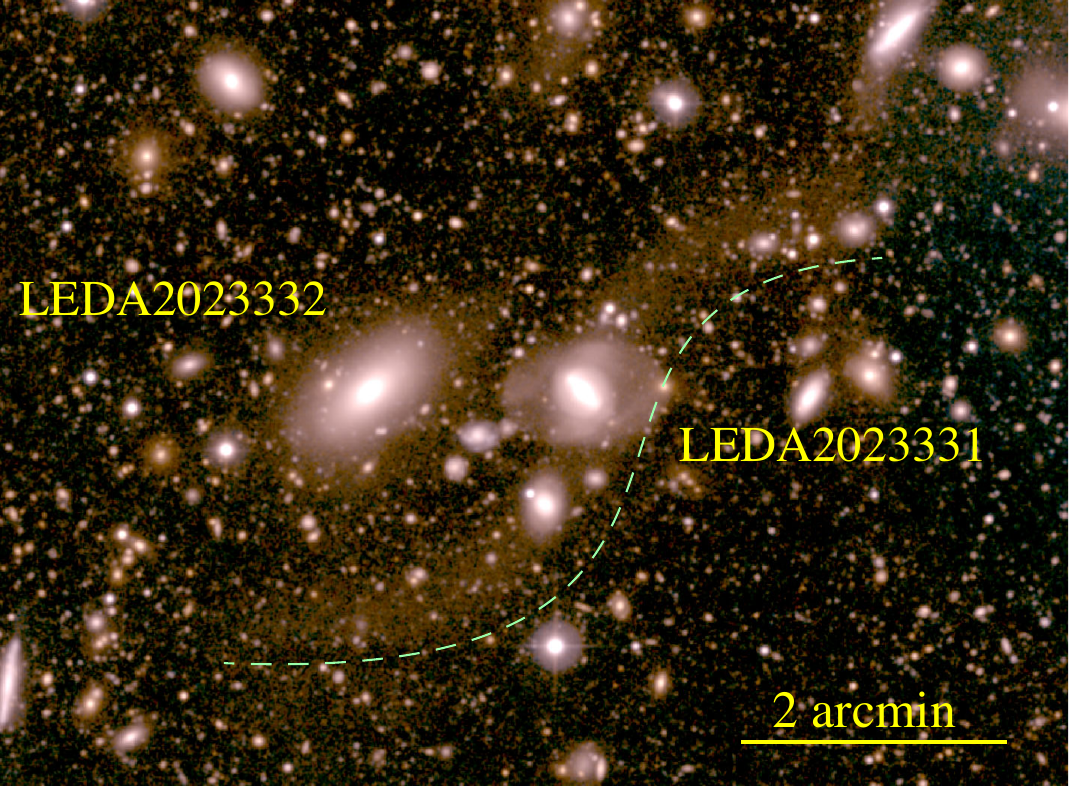}
    \caption{Region A. Tidal arm of LEDA2023331 to the left of the green dashed line.}
    \label{fig:bracos_ICL}
\end{figure}

In this region (Fig.~\ref{fig:bracos_ICL}), the galaxy LEDA2023331 shows a very extended pair of spiral arms, with a morphology suggesting a tidal effect, possibly due to an interaction with its neighbour galaxy, LEDA2023332. These galaxies are separated by 97~arcsec ($79\, h_{70}^{-1}\,$kpc) and a relative velocity of $\approx 440\,$km~s$^{-1}$ (see Table~\ref{tab:Resu12galaxies}). LEDA2023332 also shows a slightly asymmetrical light distribution.

The tidal arms of LEDA2023331 are symmetric, and each arm has an extension of approximately 150~arcsec ($\approx 120\, h_{70}^{-1}\,$kpc). The mean surface brightness of these arms is $\approx 26.3$~mag$_r$~arcsec$^{-2}$, corresponding to about $1~L_\odot\,$pc$^{-2}$.

The northern tidal arm colour is $g-r = 1.1 \pm 0.5$, redder than that of the host galaxy LEDA2023331, which has a mean colour $g-r = 0.75 \pm 0.01$. The southern arm, on the other hand, has a bluer colour, $g-r = 0.6 \pm 0.5$. However, since the tidal arms have very low surface brightness, their colour indexes have large error bars.

\subsection{Region B}

\begin{figure}[htb]
    \centering
    \includegraphics[width=0.96\columnwidth]{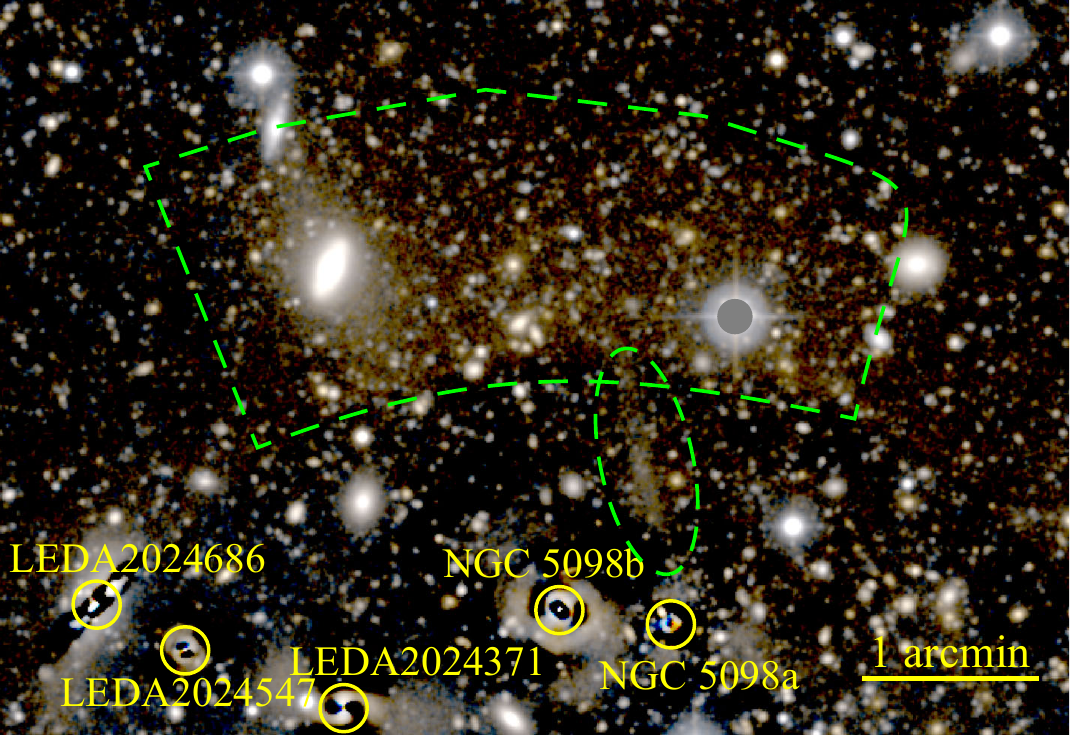}
    \caption{Region B. Intragroup light in the northern region of the NGC~5098 group (horizontal green dashed line region). A  tidal feature possibly linked to NGC~5098a (almost vertical green dashed elliptical line region) is also shown. Yellow circles show the positions of subtracted galaxies. LEDA2024686 is a foreground galaxy.}
    \label{fig:north_ICL}
\end{figure}

In this region, shown in Fig.~\ref{fig:north_ICL}, there are two faint structures revealed after subtracting the bright stars and the galaxies that we have modelled.

Towards the northern region of NGC~5098, there is a residual excess of diffuse light, an arc 1.9~arcmin ($\sim 93\, h_{70}^{-1}\,$kpc) north of the NGC~5098a and b pair, with a length of about 2.7~arcmin ($\sim 130\, h_{70}^{-1}\,$kpc). This region has a mean colour index $g-r = 1.5 \pm 0.7$, redder than the main binary pair NGC~5098a and b (that have mean $g-r$ colours 0.84 and 0.79, respectively). Since we used two S\'ersic components to model NGC~5098a, we also have an inner and outer colour index, 0.83 and 0.25 respectively, indicating a bluer extended stellar envelope.

To the north of NGC~5098a, there is a faint linear stellar stream 44~arcsec ($\sim 36\, h_{70}^{-1}\,$kpc) long (the dashed green ellipse in Fig.~\ref{fig:north_ICL}). This feature has $g-r = 1.8 \pm 0.9$, significantly redder that NGC~5098a, especially when compared to its extended stellar envelope.

\subsection{Region C}

\begin{figure}
 \centering
 \includegraphics[width=\columnwidth]{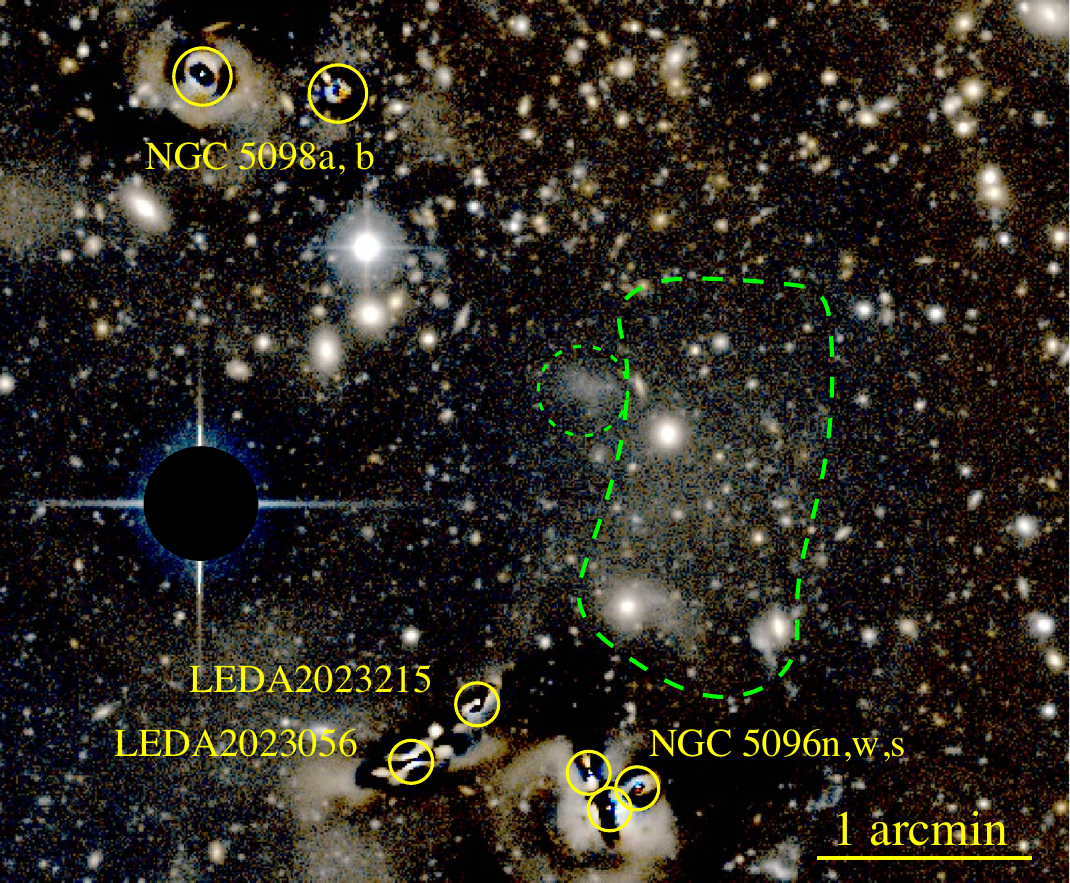}
 \caption{Region C. Diffuse intragroup light west of the line connecting NGC~5098 and NGC~5096. The green dashed contour line indicates its location. The green dotted circle shows a possible ultra diffuse galaxy (see text). Yellow circles show the positions of the subtracted bright galaxies. The centre of the star TYC 2538-952-1 is masked with a black disk with a radius of 16~arcsec.}
 \label{fig:West_ICL}
\end{figure}

In this region (see Fig.~\ref{fig:West_ICL}), we have detected a faint diffuse stellar component connecting NGC~5098 and NGC~5096 after subtracting the bright stars and galaxies. The diffuse light is roughly inside a rectangular region of $1.8 \times 0.9$~arcmin$^2$, with the larger side approximately orientated in the north--south direction. Its mean colour is $g-r = 0.2 \pm 0.3$, bluer than the main galaxies of the NGC~5098 group.

Eastward of this diffuse light component, there is a low surface brightness structure at 13$^{\rm h}20^{\rm m}09.1^{\rm s}$,  $+33^\circ07^\prime13.6^{\prime\prime}$ (J2000), with a radius of 9~arcsec, which may be an ultra diffuse galaxy. It is near the centre of Fig.~\ref{fig:West_ICL}, just to the left of the dashed green line. It is very blue, with $g-r = -0.025 \pm 0.007$.

\subsection{Region D}

\begin{figure}
    \centering
    \includegraphics[width=\columnwidth]{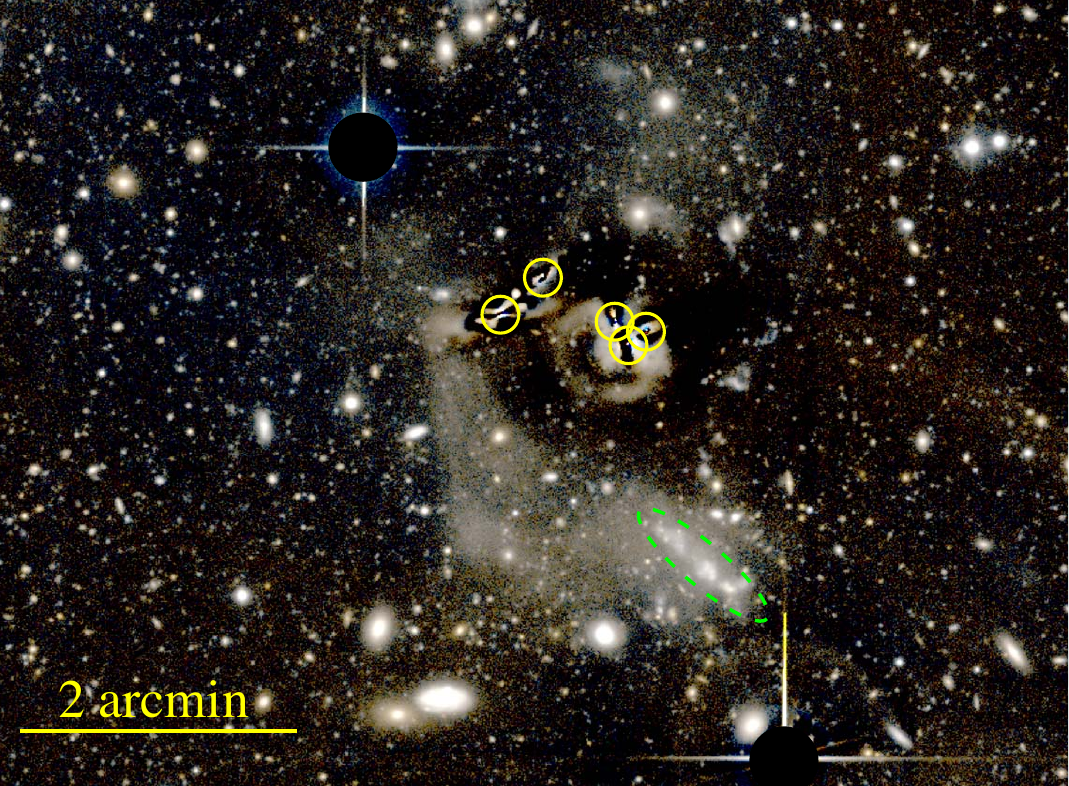}
    \caption{Region D. Image of the diffuse light residual near NGC~5096. Yellow circles indicate the position of subtracted galaxies from the NGC~5096 group. The dashed green ellipse is the galaxy WISEA J132005.90+330338.0 (see text).}
    \label{fig:SouthStructure}
\end{figure}

This region, shown in Fig.~\ref{fig:SouthStructure}, presents a bright diffuse intragroup light, extending southward from the NGC~5096 substructure. It has a blue colour, with mean $g-r =0.44 \pm 0.14$. This feature seems to originate at the western extremity of the galaxy LEDA2023056 and to end on an irregular object centred at $13^{\rm h}20^{\rm m}06^{\rm s}$,  $+33^\circ03^\prime38^{\prime\prime}$ (J2000), the green ellipse in Fig.~\ref{fig:SouthStructure}.

From the NED database, we identify this irregular object as WISEA J132005.90+330338.0 at redshift $z=0.038146$. It is a very blue galaxy, with $g-r = 0.340 \pm 0.005$, implying the presence of vigorous star formation.

At this point, we may speculate that this object could be either a tidal dwarf galaxy or the remnant of a galaxy disrupted by tidal forces. In the latter case, this could be an extreme example of a so-called jellyfish type of galaxy \citep{Ebeling2014,Poggianti2016}.

\section{Dynamical analysis}
\label{sec:Dynamic}

\subsection{Redshift Data and Radial Velocity Distribution}
\label{sec:initial_sample}

\begin{figure}
\centering
\includegraphics[width=\columnwidth]{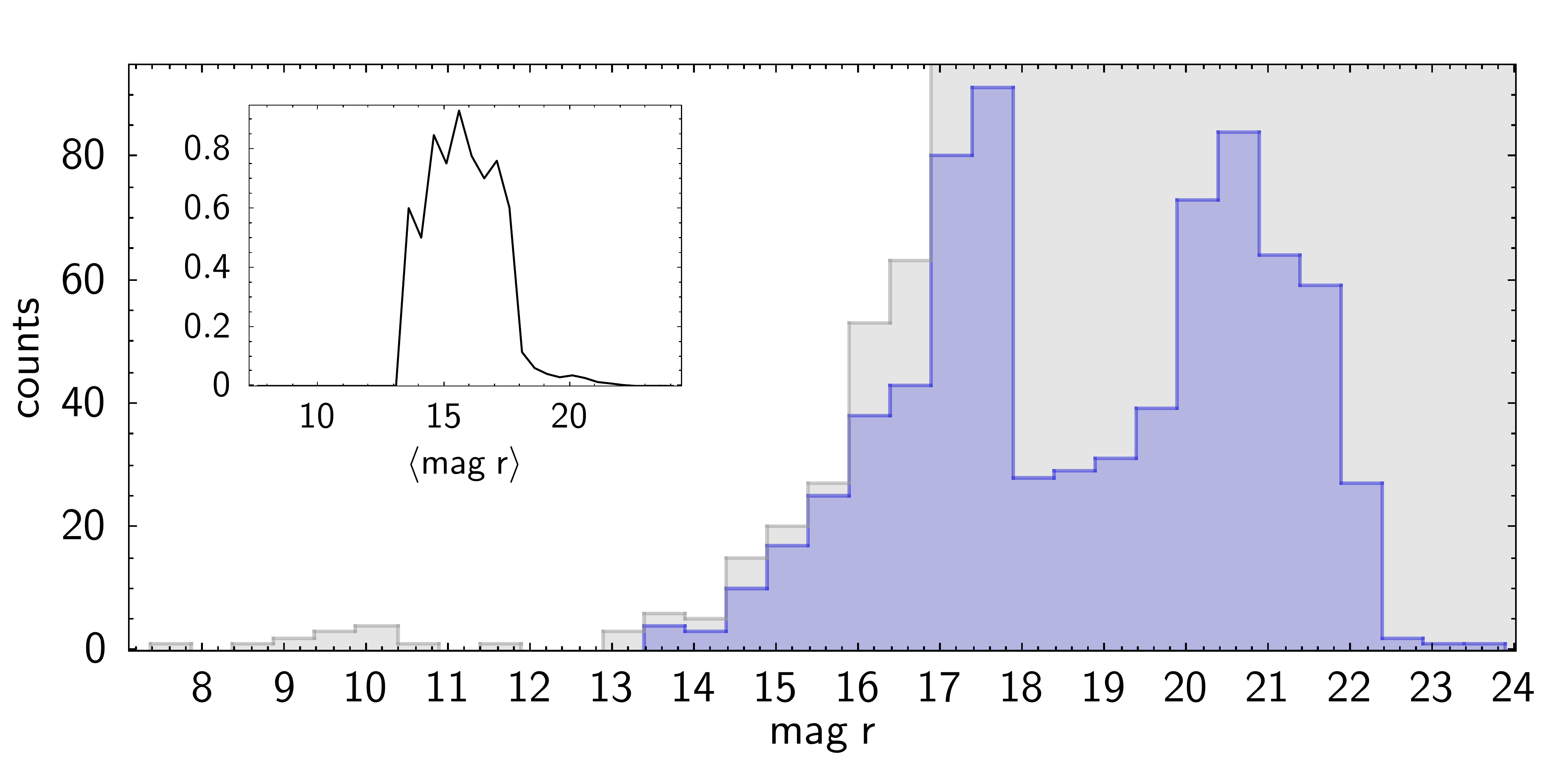}
\caption{Magnitude distribution of the redshift sample inside the 50~arcmin cone centred on NGC~5098 (\emph{blue}), compared to the magnitude distribution of the SDSS photometric sample in the same region (background \emph{light gray}). The inset shows the bin-to-bin ratio of both these magnitude distributions.}
\label{fig:FdL}
\end{figure}

The photometric and spectroscopic data were primarily extracted from the SDSS database (DR18) and complemented with redshift data 
from the Hyperleda database\footnote{\texttt{https://leda.univ-lyon1.fr/}}. We selected data in a cone with radius 50~arcmin centred on the 
fiducial position of the group  ($13^{\rm h}20^{\rm m}16.2^{\rm s}, +33^\circ 08^\prime 39^{\prime\prime}$, between the NGC~5098a+b pair). The final catalogue, comprises 748 objects with 
SDSS photometry and {with} spectroscopic redshifts $z < 1.5$. Figure \ref{fig:FdL} shows the magnitude distribution\footnote{We have used here SDSS 
dereddened magnitudes.} of galaxies in our photometric sample. As it can be seen, the completeness of our redshift sample relative to the SDSS photometric survey in the same area, is $\gtrsim 60\%$ in the magnitude range $13 < r < 18$ (see inset in Fig.~\ref{fig:FdL}).

\begin{figure}
	\centering
	\includegraphics[width=\columnwidth]{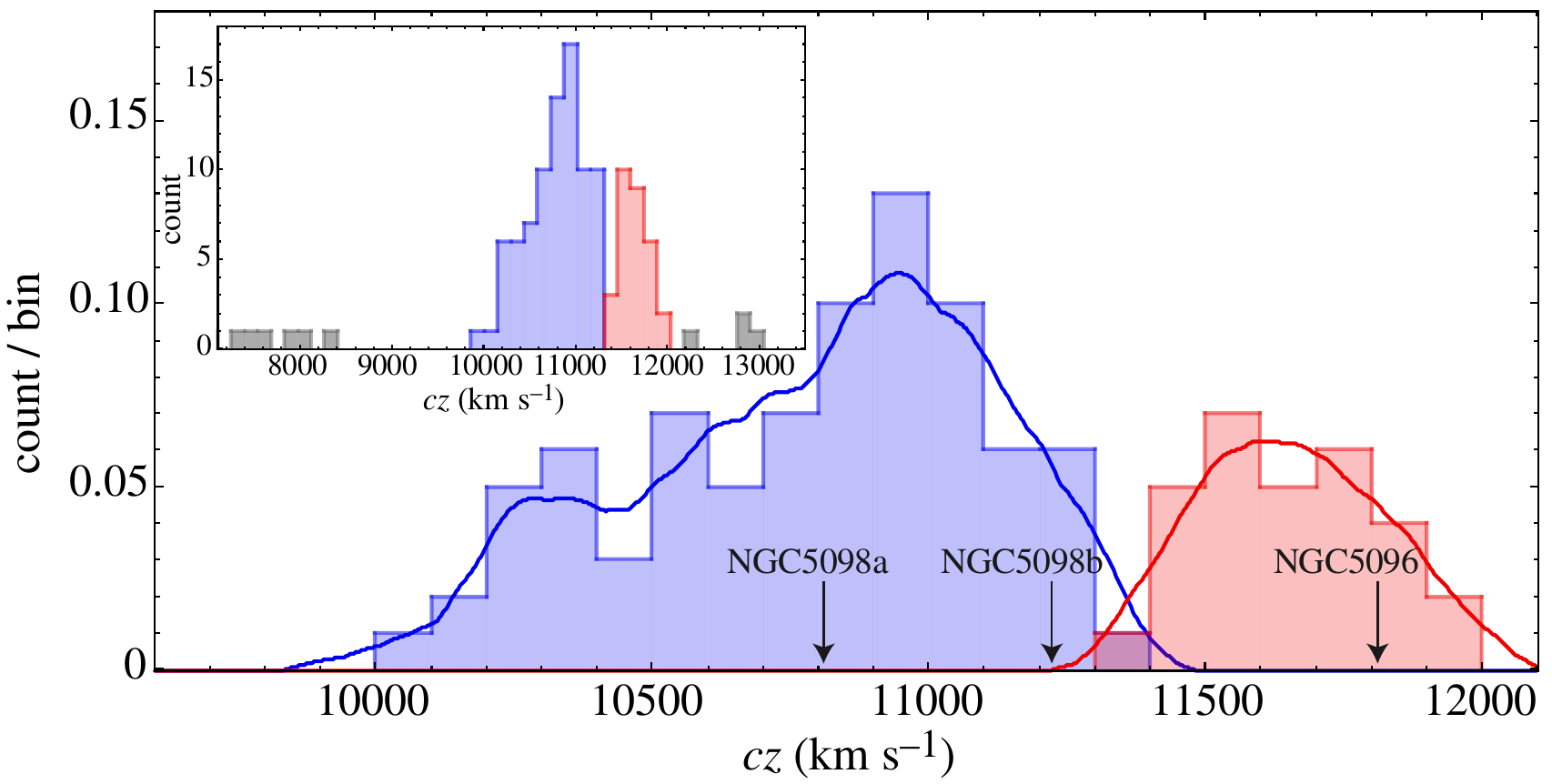}
	\caption{Velocity distributions of the two groups discussed in the text. The bins have a width of 100~km~s$^{-1}$. The arrows point the radial 
 velocities of the brightest galaxies, NGC~5098a and NGC~5098b, and NGC~5096. The continuous curves give the adaptive 
kernel approximations for the distributions. The inset displays the $0.024 < z < 0.045$ redshift distribution of 
galaxies in the 50 arcmin cone centred on NGC~5098, with velocity bins of 150~km~s$^{-1}$; grey bins indicate the galaxies that, although selected by SG, were discarded by the gap analysis.}
	\label{fig:histocz}
\end{figure}

The selection of galaxies kinematically belonging to the group was made by applying the shifting gapper technique (SG). This 
technique allows for the expected decrease in the velocity dispersion of the system with radial distance, and has proven to be 
efficient at removing outliers from the projected galaxy distribution \citep[see discussion in][]{Gifford2013}. The final list of candidate galaxies was found to 
have 116 members. Figure \ref{fig:histocz} (inset) shows the resulting radial velocity ($cz$) distribution, which is bi-modal inside the interval $10000 \la cz \la 12000\,$km~s$^{-1}$.

We analysed the bi-modality of the velocity distribution assuming that each of the two concentrations samples a normal 
population distribution. We verified this hypothesis by studying the distribution of gaps present in the data, following \citet{Beers1990}.
We define a data gap as the distance between two consecutive data values, and estimate the 
probability of its occurrence under the null hypothesis, which depends on the total sample size. We conservatively assumed 
significant deviations from normality for gaps with null hypothesis probabilities $p \leq 0.003$.

Applied to our sample, this approach strongly 
suggests a multi-modality of the underlying galaxy redshift population. It confirms the two main peaks as separate entities, suggesting that 
the four galaxies found at the extreme low and high velocities are interlopers. We end up with a list of 112 galaxy members of the group.

A similar procedure was applied by \citet{Mahdavi2005} in the analysis of their redshift sample of the NGC~5098 group.  
They found a significant gap in their (relatively small) velocity sample at $cz \sim  11400$~km~s$^{-1}$, suggesting the existence of two 
different structures in projection. This feature is visible in our sample as shown in Figure \ref{fig:histocz}. Our analysis confirms 
 the bi-modality of the velocity distribution, with one large gap ($p = 0.03$) at the same place as found by \citet{Mahdavi2005}. 
 
Figure \ref{fig:histocz} depicts with  different colours the two velocity distributions  discussed above. Notice the visually
significant negative skewness of the low velocity group (blue). However no significant large gap was found within this group.

\begin{figure}
	\centering
	\includegraphics[width=\columnwidth]{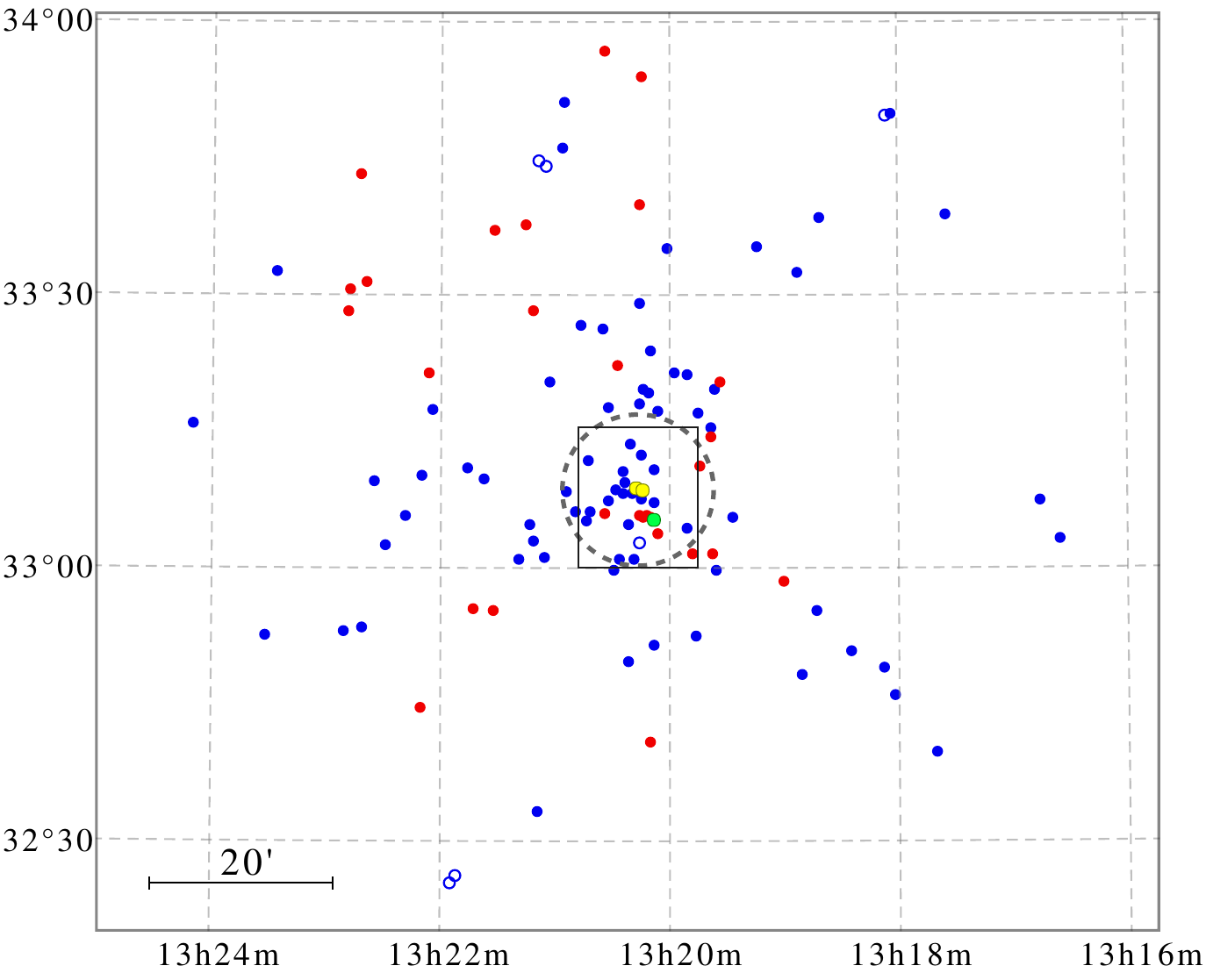}
	\caption{The sky distribution (R.A.--Decl., J2000) of galaxies according to the kinematical group they belong to. Blue circles are for group~1 (NGC~5098) galaxies and red circles for group~2 (NGC~5096) galaxies. The yellow circles show the positions of NGC~5098a and NGC~5098b, and the small green circle just below indicates the position of the NGC5096 group central galaxies. The black rectangle indicates the area covered by our CFHT images, see Fig.~\ref{fig:NGC5098_CFHT_zoomout}). The grey dashed circle is the $R_{200}$ of the NGC~5098 group (Sec.~\ref{sec:Caustics}).}
	\label{fig:sky50}
\end{figure}

\begin{figure}
   \centering
   \includegraphics[width=\columnwidth]{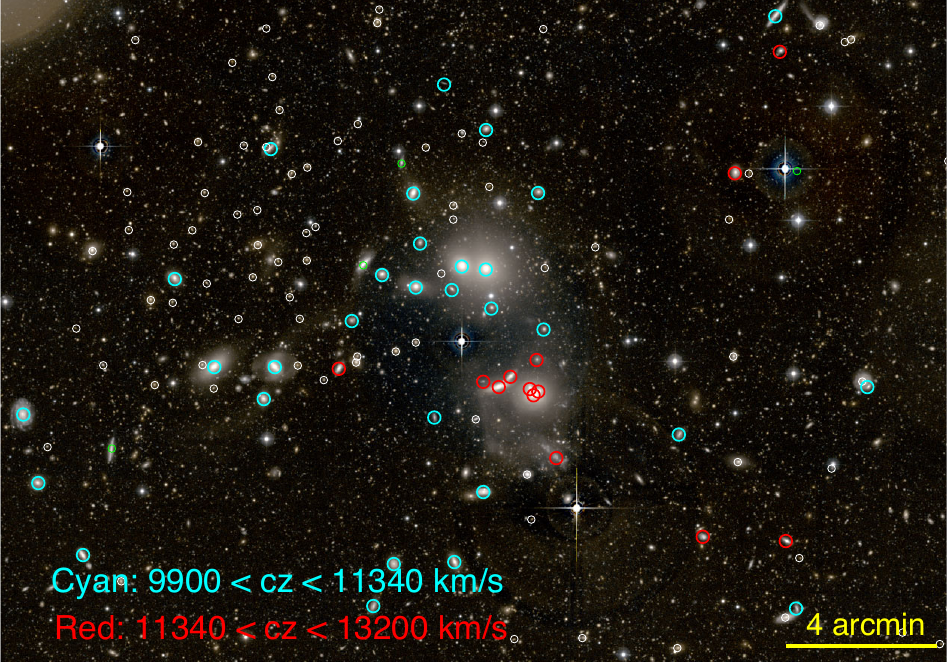}
   \caption{NGC~5098/5096 field with galaxies having a measured redshift overlaid. The cyan circles represent the galaxies that are kinematically at the location of NGC~5098, while the red circles are the galaxies kinematically near NGC~5096. Smaller gray circles are background galaxies, and a few small green circles show foreground galaxies.}
   \label{fig:cz_imagem}
\end{figure}

We conclude that the overall structure of the NGC~5098 group is dominated by two main subsystems: group 1, with $10003 \leq cz \leq 
11308$~km~s$^{-1}$, and group 2, with $11390 \leq cz \leq 11946 $~km~s$^{-1}$. The sky distribution of these groups is shown in  Figs.~\ref{fig:sky50} and \ref{fig:cz_imagem}. We see that group~1 is related to the NGC~5098 system, while group 2 is related to the NGC~5096 structure.
 
Based on the velocity distribution, group 1 has 82 members with $\langle cz \rangle = 10807^{+80}_{-81}\,$km~s$^{-1}$ and a velocity dispersion $\sigma = 318_{-44}^{+43}$~km~s$^{-1}$ (95\% confidence level). For group 2, with 30 members, $\langle cz \rangle = 11651^{+63}_{-59}\,$km~s$^{-1}$ and $\sigma = 155_{-27}^{+33}$~km~s$^{-1}$ (95\% confidence level). For the above estimates, we have assumed an error of 50~km~s$^{-1}$ on the individual line-of-sight velocity measurements.
    
\citet{Mahdavi2005} made the suggestion that these two subsystems are dynamically independent and have not yet fully 
interacted. They argue that if some interaction had already occurred, the gap between the two velocity distributions would 
not be as deep as observed, since otherwise some galaxies should already have moved to occupy that interval. However, in view of 
our discussion above, this conclusion must be somewhat modified. 

Indeed both from the point of view of the distribution of velocities (Fig.~\ref{fig:histocz}) and of their spatial distribution (Figs.~\ref{fig:sky50} and \ref{fig:cz_imagem}), the two subsystems no longer seem as isolated as suggested in the study by \citet{Mahdavi2005}, which was based on a much smaller sample. 

Furthermore, as discussed  before  (Section \ref{sec:ICL}), both the distribution of the diffuse intergalactic light and the X-ray emission \citep[e.g.][]{Randall2009} strongly  suggest that large-scale interactions may be going on within the NGC~5098 and NGC~5096 structures.

\subsection{Masses and Characteristic Radius}
\label{sec:Caustics}

The previous analysis suggests that the NGC~5098 group consists of two subsystems. One of them, subgroup 1 with 82 
members, is centred on the dominant pair of galaxies NGC~5098a-NGC~5098b, with a fairly regular distribution. Subgroup 2, 
much poorer (with 30 spectroscopic members), appears to have a spatial distribution where the majority of its galaxies lie in a line which includes the small clump of 
galaxies suggestive of a compact group, around NGC~5096, the third brightest galaxy in this region of the sky. Following 
\citet{Brinchmann2004}, both NGC~5098a and NGC~5096 are spectroscopically classified as AGN. Both systems are 
certainly interacting although maybe still in the initial phases of their future merger.

In order to better understand the dynamics of the main subgroup 1, we analysed the galaxy distribution in the projected 
phase space of the system using the method of caustics. This relies on the determination of the caustic curves 
\citep{Kaiser1987} in the projected cluster phase-space ($R, v_{\rm pec}$) as proposed by \citet{Diaferio-Geller1997}; see 
also \citet{Diaferio1999}. Here $R$ is the projected radial distance to the centre of the cluster and $v_{\rm pec}$ the line-of-sight projected peculiar velocity, i.e., the velocity referenced to the cluster mean redshift.

\begin{figure}
    \centering
    \includegraphics[width=\columnwidth]{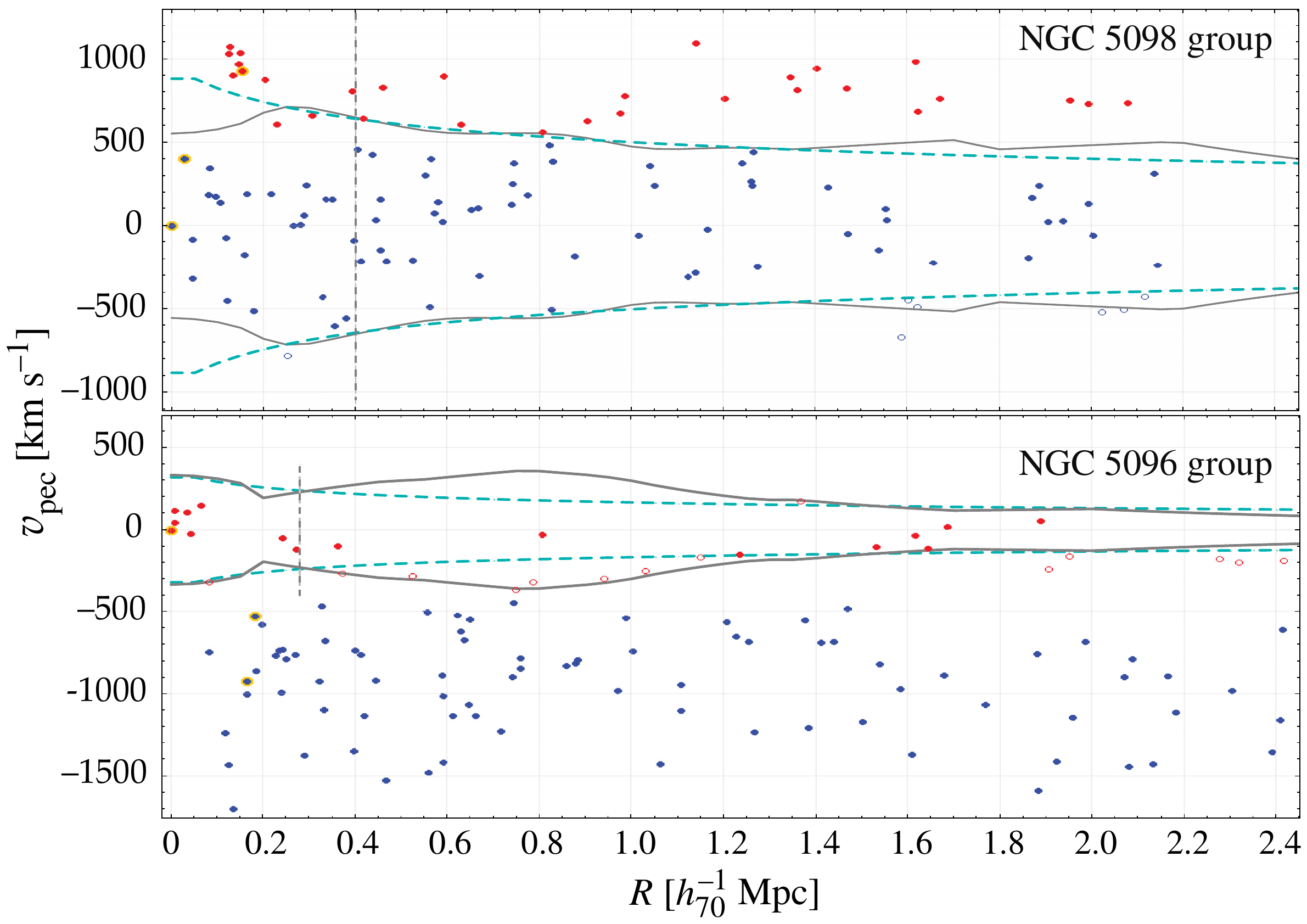}
    \caption{Projected phase space for our groups. \textsf{Top panel}: The NGC~5098 group projected phase space, $(R, v_{\rm pec})$, centred on NGC~5098a. Blue points 
    correspond to galaxies belonging to subgroup~1 and red points to subgroup~2. The dominant galaxies are marked by yellow haloes. The grey lines are the caustic lines for the subgroup~1 distribution, whereas the   green dashed lines are its best fit NFW profile, allowing for estimates of $M_{200}$ and $R_{200}$ (see text). The {\it dashed} vertical line shows $R_{200}$.  Notice the  compact  group of galaxies around NGC5096 at  $(R, v_{\rm pec}) \approx (0.1\,h_{70}^{-1}\,$Mpc,~ 1000~km~s$^{-1}$).
    \textsf{Bottom panel}: the projected phase space centred on the NGC~5096 group (subgroup 2).}
    \label{fig:VpecXR}
\end{figure}

We used the \texttt{CausticMass} code based on \citet{Gifford2013} and \citet{GiffordMiller2013}\footnote{\texttt{https://github.com/giffordw/CausticMass}}. This software was applied to the sample of 82 galaxies of subgroup 1 with the phase space centre fixed at the position of 
NGC~5098a\footnote{Changing the position of the centre $(R,V_{pec})$ does not significantly modify the final results. 
Allowing a free centre to be determined by the code itself resulted in positions very close to the central galaxy.}.  The results are shown in Fig.~\ref{fig:VpecXR} where the projected phase space positions of subgroup 2 galaxies are also displayed.

The characteristic mass and radius, $M_{200}$ and $R_{200}$,  can be obtained either from the amplitude of the caustic profile $\mathcal{A}(R)$ \citep[e.g.,][]{Diaferio1999}, or, as discussed in \citet{Gifford2017}, by fitting a NFW profile \citep{NFW96} to the caustics profile. The latter is a procedure that may result in more accurate values for $M_{200}$ and $R_{200}$. However, the NFW fitting depends very weakly on the concentration parameter, $c$, which for the NFW profile, is $c \equiv R_{200}/r_s$, where $r_s$ is the NFW scale parameter \citep[see, e.g.,][]{BivianoGerardi2003, Gifford2013}.

In order to obtain an order of magnitude of the uncertainty in the determination of the $M_{200}$ and $R_{200}$ parameters, we have run the \texttt{CausticMass} code a certain number of times, forcing the concentration parameter to assume random values sorted from a normal distribution with mean $\langle c \rangle = 5$ and dispersion  $\sigma_c =2$. These values of $c$ are adequate for $\sim 10^{13}$--$10^{14} M_\odot$ haloes \citep{Dutton2014}. Indeed, neither $M_{200}$ nor $R_{200}$ fitted values change significantly within the interval $3 \la c \la 9$.

We find for the subgroup 1 (NGC~5098), see Fig.~\ref{fig:VpecXR} top panel, 
$M_{200} = (3.93 \pm 0.28)\times10^{13}~M_{\odot}$ and
$M_{500} = (2.55 \pm 0.32)\times10^{13}~M_{\odot}$. 
For the radii,
$R_{200}= (0.69 \pm 0.02)\,h_{70}^{-1}\,$Mpc and
$R_{500}= (0.44 \pm 0.02)\,h_{70}^{-1}\,$Mpc.

These values are consistent with the mass determination by \citet{Sun2009}, based on \textit{Chandra} data, $M_{500} = 2.00^{+0.28}_{-0.48} \times 10^{13} M_\odot$, where their value of $R_{500}$ is $392^{+17}_{-32} h_{70}^{-1}\,$kpc (about 8.2~arcmin). Finally, applying the caustic method, we could derive a velocity dispersion $\sigma \simeq 270$~km~s$^{-1}$ for the main subgroup 1.

For the second subgroup, centred on NGC~5096, (Fig.~\ref{fig:VpecXR}, bottom panel) we find 
$M_{200} = (0.17 \pm 0.02) \times 10^{13}~M_{\odot}$ and 
$M_{500} = (0.13 \pm 0.03) \times 10^{13}~M_{\odot}$.
For the radii we have
$R_{200} = (0.24 \pm 0.01) h^{-1}_{70}\,$Mpc and
$R_{500} = (0.17 \pm 0.01) h^{-1}_{70}\,$Mpc.
The corresponding velocity dispersion is $\sigma \simeq  90$~km~s$^{-1}$ for the central galaxies of the NGC~5096 compact group.

\section{Discussion}
\label{sec:Discussion}

Both photometric and kinematic analyses show, at least, a bimodal structure, as already suggested in the literature \citep{Mahdavi2005}.

New deep imaging shows several low surface brightness structures which may be the result of stellar stripping by tidal forces, either due to the encounter of galaxies, or to interactions during the passage of a substructure through the main group of NGC~5098. Some features, such as the tidal arms shown in Fig.~\ref{fig:bracos_ICL} (Region A) probably result from two-body interactions of galaxies near the core of the NGC~5098 group. Given the large error bars, the tidal arm colours are roughly the same as the colour of LEDA2023331, although the southern arm seem to be bluer.

The bright diffuse stellar component seen in Region D, Fig.~\ref{fig:SouthStructure}, is possibly the result of a tidal interaction of a gas rich, blue galaxy, probably WISEA J132005.90+330338.0, with the core of the compact group around NGC~5096. The hypothesis that it is a gas rich object stems from the blue colour, $g-r = 0.34$. of its stellar population.

The low surface brightness features in Region B (Fig.~\ref{fig:north_ICL}) may indicate some older tidal stripping, since they are quite red (notwithstanding the large error bars). It is not clear which pairs of galaxies were involved in the formation of these diffuse components. However, given their location, both structures may be related to the central galaxy NGC~5098a.

We have detected a very faint diffuse stellar emission between the core of the NGC~5098 group and the compact substructure of NGC~5096 (Region D, Fig.~\ref{fig:West_ICL}). It has a bluer colour ($g-r = 0.44$) compared with the NGC~5098a and NGC~5098b pair, and also compared to the compact core of NGC~5096. There is no galaxy clearly linked to this structure, but its location suggests that it may have been produced by a  past interaction between one or more galaxies pertaining to the NGC~5098 group and to the NGC~5096 compact group, assuming it has passed near the central region of the NGC~5098 group. The large-scale view of the NGC~5098 and NGC~5096 structure, displayed in Figure~\ref{fig:NGC5098_r_semStar_masked}, shows a large diffuse intragroup light component linking both substructures. The blue colour of this diffuse component suggests it originates from the tidal stripping of galaxies \citep[see, e.g.][]{DeMaio2018,Contini2019,Tang2023}.

This scenario, where NGC~5096 has passed near the core of the NGC~5098 group, is further corroborated by our dynamical analysis. We have clearly detected two structures in velocity space, separated by $\approx 700\,$km~s$^{-1}$. The gap analysis done with a sample of 112 galaxies suggests that the groups are not isolated and may have undergone recent interactions.

Further evidence for the past interaction between the groups comes from the X-ray analysis by \citet{Randall2009}, where they detect a spiral arm feature which is typically related to a sloshing phenomenon \citep[e.g.,][]{Ascasibar2006,Machado2015}. In Fig.~\ref{fig:Chandra_contGalaxies} we show that the diffuse X-ray emission has a secondary component around NGC~5096, linked to the main emission from the NGC~5098 group.

While  \citet{Randall2009} suggest that the sloshing is due to the arrival of NGC~5098b, stripped from its own hot gas, and perturbing the gas around NGC~5098a, we tentatively suggest that the sloshing may be due to the passage of the core of the NGC~5096 group near the centre of the NGC~5098 group. This hypothesis remains to be tested by means of simulations.

Dedicated hydrodynamical $N$-body simulations would be needed in order to carefully evaluate whether NGC~5096 is a plausible candidate to be the perturber that induced the sloshing in NGC~5098. In particular, the temperature, morphology and orientation of the sloshing spiral must be consistent with the current location of the presumed perturber. Even in the scenario where the two clusters have already interacted recently, one cannot confidently rule out the possibility that the sloshing in NGC~5098 may have been induced by an earlier encounter with a separate perturber.

Alternatively, it is conceivable that the two groups might be incoming for a first approach. If they have not closely interacted yet, then the observed intragroup light could be due to tidal stripping of galaxies within each group, separately. In that case, the sloshing seen in X-rays in NGC 5098 would have been triggered by some other perturber, yet to be identified.

\section{Conclusions}
\label{sec:Conclusion}

We have analysed here deep imaging in the $g$ and $r$ bands acquired with CFHT/MegaCam and the available redshift measurements within a radius of 50~arcmin around the NGC~5098 and NGC~5096 groups. We also used publicly available data from a \textit{Chandra} observation.

We confirm that NGC~5098 and NGC~5096 are two dynamically distinct groups, with a velocity difference about 700~km~s$^{-1}$ and we suggest that they are already interacting, with a possible previous passage of NGC~5096 near the core of NGC~5098. This conclusion comes from the intragroup light that we detect, which may be due to tidal stripping from galaxy members of both groups. We posit that this interaction may be the reason for the sloshing feature observed in X-rays, and for the extended emission linking both groups. The skewness observed in the velocity distribution of the NGC~5098 group may also be a result of the perturbation induced by the passage of the NGC~5096 group. 

Further work could greatly benefit from numerically modelling this complex system with customized hydrodynamical simulations.

%%%%%%%%%%%%%%%%%%%%%%%%%%%%%%%%%%%%%%%%%%%%%%%%%% from MNRAS, could be useful...
\section*{Data Availability}

Reduced optical images in $g$ and $r$ bands of NGC~5098/5096 are available in electronic form at the CDS via anonymous ftp to cdsarc.u-strasbg.fr (130.79.128.5).

%%%%%%%%%%%%%%%%%%%%%%%%%%%%%%%%%%%%%%%%%%%%%%%%%
\begin{acknowledgements}
We thank the referee, Dr. Emanuele Contini for constructive suggestions that helped improve this paper. G.B.L.N. thanks the hospitality of IAP where part of this work was done and is grateful for the financial support from CNPq under grant 314528/2023-7, FAPESP under grant 2024/06400-5, and CNRS. F.D. is grateful to CNES for financial support. REGM acknowledges support from CNPq, through grants 406908/2018-4 and 307205/2021-5, and from \textit{Fundação Araucária} through grant PDI 346/2024 -- NAPI \textit{Fenômenos Extremos do Universo}.
 
\end{acknowledgements}


\bibliographystyle{aa}
\bibliography{example} % if your bibtex file is called example.bib

\end{document}